\documentclass[nofootinbib,twocolumn,preprintnumbers,amsmath,showkeys,amssymb,showpacs]{revtex4}
\def\JMP#1#2#3{J. Math. Phys. {\bf #1}, #2 (19#3)}
\def\CQG#1#2#3{Class. Quantum Grav. {\bf #1}, #2 (19#3)}
\def\PLB#1#2#3{Phys. Lett. B {\bf #1}, #2 (19#3)}
\def\PRD#1#2#3{Phys. Rev. D {\bf #1}, #2 (19#3)}

\begin{document}
\preprint{IPM/P-2008/063 \cr
	 \eprint{arXiv:0811.4238v1 [gr-qc]}}

\title{Effectively Emergent Quantum Mechanics}

\author{Qasem \surname{Exirifard}}
\affiliation{
School of Physics, Institute for Research in Fundamental Sciences, P.O.Box 19395-5531, Tehran, Iran}

\email{exir@theory.ipm.ac.ir}

\begin{abstract}
We consider non minimal coupling between matters and gravity in modified theories of gravity.  In contrary to the current common sense, we report that quantum mechanics can effectively emerge when the space-time geometry is sufficiently flat. In other words, quantum mechanics  might play no role when and where the space-time geometry is highly curved.

We study the first two simple models of Effectively Emergent Quantum Mechanics(EEQM): R-dependent EEQM and G-dependent EEQM where R is the Ricci scalar and G is the Gauss-Bonnet Lagrangian density. We discuss that these EEQM theories might be fine tuned to remain consistent with all the implemented experiments and performed observations. In particular,  we observe that G-dependent EEQM softens the problem of quantum gravity.
\end{abstract}

\keywords{ modified theories of gravity, quantum mechanics, quantum gravity, modified uncertainty relation}
\pacs{04.50.Kd,04.60.m,03.65.w}
\maketitle

We have not yet constructed a theory of Quantum Gravity despite of various attempts within the paradigm of the string theory, the paradigm of  canonical and loop quantum gravity,   and so on. Frankly speaking we are still facing  the problem of quantum gravity. This problem is Platonic and theoretical in the sense that we have not yet empirically and directly explored the regimes wherein quantum gravity is supposedly important.\footnote{The nature also appears not to have the tendency to let us put our hands where quantum gravity is supposedly important: Recall the paradigm of the cosmological inflation, and the presence of the event horizons cloaking the singularities of the black hole in the center of galaxies} In contrary to our current common sense, here, we are going to explore the possibility that maybe quantum mechanics emerges when the space-time geometry is sufficiently flat. In other words, we are going to consider the possibility that quantum mechanical corrections are highly suppressed when or where the space-time geometry is highly curved.

Quantum mechanics states that momentum and position of no particle can be simultaneously measured:
\begin{equation}\label{Hizenberg}
 \Delta x \Delta p \geq \frac{\hbar}{4}\,, 
\end{equation}
where $\Delta x$ and $\Delta p$ are respectively the uncertainty is the position and momentum of the particle. When we perform a measurement on a particle, the particle is localized where we are doing the experiment. So some amount of information about the space-time geometry might be already hidden in \eqref{Hizenberg}. In other words, the uncertainty relation can read
\begin{equation}\label{HizenbergG}
 \Delta x \Delta p \geq \frac{\hbar}{4}\,F(R_{\mu\nu\lambda\eta},\nabla_\eta,g_{\mu\nu}), 
\end{equation}
where $F$, being a number, is a functional of the Riemann tensor ($R_{\mu\nu\lambda\eta}$), the covariant derivative ($\nabla_\eta$) and the metric ($g_{\mu\nu}$) of the space-time inside the laboratory. We notice that $F(R_{\mu\nu\lambda\eta},\nabla_\eta,g_{\mu\nu})\sim 1$ must hold for the space-time geometries wherein we have directly or indirectly performed some experiments. $F(R_{\mu\nu\lambda\eta},\nabla_\eta,g_{\mu\nu})$, however, can approaches zero when and where the space-time geometry becomes sufficiently curved.  Therefore there exists the possibility that the uncertainty relation approaches $\Delta x \Delta p \geq 0$ as we move backward in time toward the moment of the big bang. So Quantum Mechanics may not effectively exist when the universe was formed. Quantum Mechanics can effectively emerge when the universe is old enough. We refer to this  possibility as Effectively Emergent Quantum Mechanics, or in brief EEQM. We discuss that EEQM  naturally rise in modified theories of gravity when  matters  non-minimally couple to gravity.

\section{R-dependent EEQM}
The simplest form for  $F(R_{\mu\nu\lambda\eta},\nabla_\eta,g_{\mu\nu})$ would be  $F(R_{\mu\nu\lambda\eta},\nabla_\eta,g_{\mu\nu})=f(R)$. In the R-dependent EEQM the uncertainty relation reads
\begin{equation}\label{HizenbergR}
 \Delta x \Delta p \geq \frac{\hbar}{4}\,f(R), 
\end{equation}
where we should require $f(R)=1$ for $R \in [ \Lambda, 10^{40} \Lambda]$ where $\Lambda=10^{-120}\frac{1}{l_{planck}^2}$ is the cosmological constant. Note that $R=10^{40} \Lambda$ is at the order of magnitude of the Ricci scalar inside the Sun. But what kind of theories could effectively give rise to  eq. \eqref{HizenbergR}? In order to answer this question let us briefly review quantum field theory. 

Loosely speaking QFT means computing the path integral. The path integral of the Einstein-Hilbert gravity reads
\begin{equation}\label{Z}
\boldsymbol{Z}\,=\, \int D\Psi Dg e^{-i \int d^4 x \frac{\sqrt{-\det g}}{\hbar}(-8 \pi G R + L_{\text{Matters}}(\Psi))}\,.
\end{equation}
The path integral of a modified gravity given by
\begin{equation}
S= \int d^4x \sqrt{-\det g} \frac{1}{f(R)}( - 8 \pi G R + L_{\text{Matters}}(\Psi))\,,
\end{equation}
reads
\begin{equation}\label{MOZ}
\boldsymbol{Z}\,=\, \int D\Psi Dg e^{-i \int d^4 x \frac{\sqrt{-\det g}}{f(R)\hbar}(-8 \pi G R + L_{\text{Matters}}(\Psi))}\,.
\end{equation}
By comparing \eqref{Z} and \eqref{MOZ}, we conclude that when the space-time is sufficiently homogeneous than the effective Hisenberg constant for \eqref{MOZ} is $f(R) \hbar$. Therefore if we choose $f(R)$ such that 
\begin{equation}
\lim_{R\to \infty} f(R)\,=\,0\,,
\end{equation}
then the quantum fluctuation of matters turn negligible at the moment of big bang where  $R$ is very large.  Since the effective Planck constant reads: 
\begin{equation}
l_{eff-planck}^2\,=\, \frac{G \hbar_{eff}}{c^3}\,=\, \frac{G \hbar}{c^3} f(R)\,,
\end{equation}
then the perturbative quantum corrections to the effective gravity reads
\begin{equation}
L_{gravity}\,=\, \frac{R + O(l_{eff-planck}^2 R^2)}{f(R)} \,=\,\frac{ R }{f(R)} [1+ O(R f(R))]\,.
\end{equation}
So if we also choose $f(R)$ such that 
\begin{equation}
\lim_{R\to \infty} R\, f(R)\,=\,0
\end{equation}
then the quantum corrections to the gravity will be negligible. 

A simple form $f(R)$ that satisfy the above criteria is
\begin{equation}
f(R)= \tanh^2(\frac{\alpha}{R})\,,
\end{equation}
where $\alpha$ is constat value sufficiently larger than $10^{40} \Lambda$.  But what would be the value of $\alpha$?  A natural choice for $\alpha$ is to choose  a value compatible with in the inflationary paradigm \cite{Exirifard:2008iy}: $\alpha\sim 10^{14} Gev$.  Note that inflation naturally can occur in EEQM.\footnote{In particular, R-dependent EEQM is a F(R) gravity. F(R) gravity are known to have an equivalent description in Enisten-Hilbert action coupled to an scalar field \cite{E1,E2,E3,E4,E5} (or look at eq. 4-6 of \cite{chiba}). This scalar field can be inflaton. A set of $F(R)$ gravities can naturally describe inflation.}  In this standpoint,  however, inflation is a transition from  `classical' world toward where quantum mechanics is significant as much as today is.

\section{$\boldsymbol{G}$-dependent EEQM}
The obvious problem of R-dependent EEQM is that after the inflation, we still will need the quantum theory for gravity  around small black holes, black holes that  outside of horizon of which $R_{ijkl}R^{ijkl}\sim \frac{1}{l_{planck}^4}$. If we consider other types of EEQM wherein the effective $\hbar$ depends on other scalars rather than the Ricci scalar, we then can make quantum mechanical effects irrelevant also in the vicinity of what replace small black holes in those theories. To  this end, there exist infinite possibilities. The anomalous flat rotational curves of the spiral galaxies suggests that we should choose $\frac{|R_{ijkl}|^4}{|\nabla_k R_{ijkl}|^2}$ \cite{Exirifard:2008dy}. Here, we however consider the simplest choice: $\boldsymbol{G}$-dependent EEQM, 
\begin{equation}\label{HG}
 \Delta x \Delta p \geq \frac{\hbar}{4}\,f(\boldsymbol{G}), 
\end{equation}
where $\boldsymbol{G}$ stands for the Gauss-Bonnect Lagrangian density in the laboratory.  A modified gravity that leads to \eqref{HG} (around flat space-time geometry) reads
\begin{equation}\label{fG-gravity}
S= \int d^4x \sqrt{-\det g} \frac{1}{f(\boldsymbol{G})}( - 8 \pi G R + L_{\text{Matters}}(\Psi)),
\end{equation}
where we should assume $f(\boldsymbol{G})\sim 1$ for $\sqrt{\boldsymbol{G}} \in [\Lambda-10^{40}\Lambda]$. If we choose $f(\boldsymbol{G})$ such that
\begin{eqnarray}
\lim_{\boldsymbol{G}\to \infty} f(\boldsymbol{G})&=&0\,,\\
\lim_{\boldsymbol{G}\to \infty} \boldsymbol{G} f(\boldsymbol{G})&=&0\,,
\end{eqnarray}
 then the perturbative quantum corrections to \eqref{fG-gravity} remain sufficiently small  for all values of $\boldsymbol{G}$. An example of $f(\boldsymbol{G})$ that meets all these constraint is
\begin{equation}
f(\boldsymbol{G})= \tanh^2(\frac{\alpha^2}{\boldsymbol{G}})\,,
\end{equation}
 where for $\alpha\sim 10^{14} Gev$, we might encode inflation in \eqref{fG-gravity}. We notice that the above conditions on $f(\boldsymbol{G})$ can exist  in addition to cosmologically viable constraints of `$f(\boldsymbol{G})$'  gravity \cite{DeFelice:2008wz}. Therefore,  the exact solutions of \eqref{HG} and their properties will remain intact when we add quantum corrections. When a black hole is large, the Schwarzschild black-hole remains a perturbative solution of \eqref{fG-gravity}. But when  mass is small we need to obtain the exact solution of \eqref{fG-gravity}.  Once we find the exact solutions, we know that quantum corrections do not significantly change them. 

\section{Conclusions and outlook}
Starting from a covariantly modified uncertainty relation, we have realized a set of modified gravities wherein matters non-minimally couple to gravity, and quantum corrections are suppressed where or when the space-time geometry is highly curved. In other words,  quantum mechanics  plays no role when and where the space-time geometry is highly curved.

\end{document}